\documentclass{article}
\usepackage{spconf,amsmath,amssymb,graphicx}

\usepackage{cite}
\usepackage{algorithmic}
\usepackage{graphicx}
\usepackage{textcomp}
\usepackage{xcolor}
\usepackage{booktabs}
\usepackage{multirow}
\usepackage{multicol}
\usepackage{stfloats}

\usepackage{subcaption}
\renewcommand{\thetable}{\arabic{table}} % Use Arabic numerals
\usepackage{hyperref}
\hypersetup{
    colorlinks=true,
    linkcolor=blue,
    filecolor=magenta,      
    urlcolor=blue,
    citecolor = blue,
}

\def\x{{\boldsymbol{x}}}
\def\y{{\boldsymbol{y}}}
\def\z{{\boldsymbol{z}}}

\newcommand{\bemc}[1]{{\color{black}{#1}}}

\title{Scale Equivariance Regularization and Feature Lifting \\in High Dynamic Range Modulo Imaging}
\name{Brayan Monroy and Jorge Bacca \vspace{-0.7em}} 
\address{Department of Computer Science, Universidad Industrial de Santander, Colombia.}
\begin{document}
%\ninept
%
\maketitle
\begin{abstract}
Modulo imaging enables high dynamic range (HDR) acquisition by cyclically wrapping saturated intensities, but accurate reconstruction remains challenging due to ambiguities between natural image edges and artificial wrap discontinuities. This work proposes a learning-based HDR restoration framework that incorporates two key strategies: (i) a scale-equivariant regularization that enforces consistency under exposure variations, and (ii) a feature lifting input design combining the raw modulo image, wrapped finite differences, and a closed-form initialization. Together, these components enhance the network's ability to distinguish true structure from wrapping artifacts, yielding state-of-the-art performance across perceptual and linear HDR quality metrics.
\end{abstract}
\section{Introduction} \vspace{-0.2em}
\label{sec:intro}

High dynamic range (HDR) imaging plays a crucial role in the capture and rendering of scenes that exhibit significant variations in luminance levels, as commonly found in real‑world environments~\cite{reinhard2021high,contreras2025high}. Conventional imaging sensors, including charge-coupled device (CCD) and Complementary Metal-Oxide-Semiconductor (CMOS) devices, are limited by their well capacity and quantization precision, resulting in dynamic range constraints that often lead to signal clipping in overexposed regions~\cite{bandoh2010recent}. This saturation effect causes the loss of critical scene details in bright areas while simultaneously under-representing information in darker regions. To address this limitation, several computational and hardware‑based solutions have been proposed, such as multi-exposure fusion and bit-depth enhancements~\cite{debevec2023recovering, robertson2003estimation}; however, these approaches often incur trade‑offs involving spatial resolution or hardware complexity, which restrict their practical deployment of dynamic scenes or resource‑constrained settings.

Modulo imaging has recently emerged as a promising alternative to conventional HDR acquisition strategies~\cite{zhao2015unbounded, bhandari2017unlimited}. In these approaches, the intensity of light is cyclically wrapped at pixel level once they exceed a predefined threshold, thereby preventing saturation and capturing signals that extend beyond the native dynamic range of the sensor~\cite{zhao2015unbounded, bhandari2020hdr}. The resulting modulo images must then be processed to unwrap/reconstruct the underlying HDR scene~\cite{bhandari2020unlimited}. A variety of restoration algorithms have been developed for this inverse problem. For example, \bemc{ PnP-UA mitigates wrap artifacts through denoising regularization~\cite{bacca2024deep}}, AHFD, based on unlimited sampling, uses stripe artifact estimation to remove artifacts~\cite{monroy2025autoregressive}, and UnModNet employs an image edge predictor coupled with level wrap estimation~\cite{zhou2020unmodnet}; however, these alternatives struggle in high lighting conditions.

In parallel, Equivariant Imaging (EI) is a framework for image restoration in inverse problems by exploiting known invariance of the signal set to guide reconstruction~\cite{chen2021equivariant}. In this paradigm, one enforces that the composition of the reconstruction network with the sensing model is equivariant with a chosen group of transformations, thus constraining the network to recover components that lie in the null space of the forward operator~\cite{chen2023imaging, schwab2019deep}. By integrating reconstruction consistency with an equivariant loss, networks can learn to infer missing information~\cite{tachella2023sensing,sechaud2024equivariance, monroy2025generalized,bacca2025projection}.

% In this work, we propose an equivariant regularization of a modulo imaging restoration network under the scaling operator. Since the same HDR scene with different exposure times will produce different modulo images, the output of an ideal restoration network should be the same HDR estimation scaling by a factor. By incorporating these properties into model training by loss regularization, we can effectively enable the restoration model to distinguish between the edges of natural images and modulo wraps. In addition, we explore the possibility to guide the restoration task by evaluating different input alternatives, from modulo image, first finite wrapped difference, and initial guess of a close-form solution of measurement consistency loss.

In this work, we propose an equivariant regularization of a modulo imaging restoration network under the scaling operator. Varying exposure times for the same HDR scene produce different modulo images, thus an optimal restoration network should deliver appropriately scaled HDR estimates. These properties are integrated into the model training via loss regularization, enhancing model generalization. In addition, we explore how to guide the restoration task through `feature lifting' of the network input. Specifically, we evaluate the influence of three features: the modulo image, the modulo differences, and a closed-form initialization derived from the measurement consistency loss. Simulation results \bemc{show that the proposed method outperforms} state-of-the-art methods by up to 4dB in PSNR-Y and 2dB in PSNR in PU21 encoding.

\vspace{-0.8em}

% \section{Modulo Imaging}
% \label{sec:format}

% Modulo imaging is an emerging sensing strategy designed to extend the dynamic range of conventional imaging sensors. In a modulo imaging system, the sensor performs a cyclic reset of pixel intensities whenever they exceed a predetermined threshold. This process allows the sensor to capture scene radiance beyond the native well capacity without saturation. The captured image thus contains wrapped or folded intensities that require post-processing to reconstruct the underlying high dynamic range (HDR) scene.

% Let $\x \in \mathbb{R}^{m\times n}$ represent the true HDR image, and let $\y \in \mathbb{R}^{m \times n}$ denote the modulo measurement. The image formation model can be described as
% \begin{equation}
%     \y = \mathcal{W}_b(\x) = \operatorname{mod}(\x, 2^b) = \x - 2^b \cdot \Big\lfloor \frac{\x}{2^b} \Big\rfloor ,
% \end{equation}
% where $\mathcal{W}_b(\cdot)$ is the modulo operator with a reset threshold of $2^b$, corresponding to a $b$-bit sensor. 

% In the absence of noise, the relation between the true image $x$ and the modulo measurement $y$ can be written as
% \begin{equation}
%     \x = \y + \k \cdot 2^b,
% \end{equation}
% where $\k \in \mathbb{Z}^{m \times n}$ is an integer-valued matrix indicating the number of resets for each pixel.

\section{Modulo Imaging} \vspace{-0.2em}
\label{sec:format}

Modulo imaging extends the dynamic range of a $b$ bit sensor by performing a cyclic reset of each pixel when its intensity exceeds the well capacity threshold $2^b$. Concretely, let $\x\in\mathbb{R}^{m\times n}$ be the true HDR image; the sensor records the \bemc{`modulo'} measurement as
\[
\y \;=\;\mathcal{W}_b(\x)
\;=\;\operatorname{mod}(\x,2^b)
\;=\;\x \;-\;2^b\,\Big\lfloor\frac{\x}{2^b}\Big\rfloor,
\]
where $\mathcal{W}_b(\cdot)$ is the modulo operator with reset threshold $2^b$, and $\lfloor\cdot\rfloor$ denotes the element‑wise floor function.

% To address the reconstruction problem linearly, consider that for reconstruction of $\x$ from $\y$, one assumes that the finite differences of $\x$, $\Delta \x$, remain consistent within the modulo domain, excluding wrapping discontinuities. The objective is then to reformulate the problem to minimize the discrepancy between the spatial differences of $\x$ and the modulo adjusted or rewrapped differences of $\y$, while including a regularization term reflecting prior knowledge.
% \begin{equation}
%     \underset{\x}{\text{minimize}} \;  \left\| \Delta \x - \mathcal{M}_{b}(\Delta \y) \right\|_2^2 + \lambda g(\x),
% \end{equation}

To obtain a tractable linear reconstruction of the HDR image $\x$ from its modulo measurements $\y$, the problem is formulated as a 2D unwrapping problem under the {\it Itoh condition}~\cite{itoh1982analysis}.  Define the finite difference at the pixel $(i,j)$ by $\Delta x_{i,j}=[x_{i+1,j}-x_{i,j},\,x_{i,j+1}-x_{i,j}]$, and similarly $\Delta \y$. As long as $|\Delta \x |<2^{b-1}$, the modulo finite differences  
\[
\mathcal{M}_b(\Delta \y ) 
=\Delta \y -2^b \text{round} \bigl(\tfrac{\Delta \y  }{2^b}\bigr),
\]
coincide with $\Delta \x$. Then, the optimization problem seeks to recover $\x$
 by reducing the gap between the spatial differences of $\x$ and
 the modulo finite difference of $\y$ as follows
\[
\hat{\x}=\arg\min_{\x}\|\Delta \x-\mathcal{M}_b(\Delta \y)\|_2^2+\lambda\,g(\x),
\]
where $g(\x)$ encodes prior knowledge (e.g.\ smoothness or sparsity) and $\lambda>0$ balances fidelity versus regularization.

% \begin{figure*}
%     \centering
%     \includegraphics[width=\linewidth]{visual_results.png}
%     \caption{\textbf{Visual Results of HDR Recovery Methods.} Qualitative comparison of unwrapped images from various methods, demonstrating the proposed approach achieves superior visual fidelity and quantitative metrics (PSNR-Y|SSIM|MSSSIM) on PU21 encoding and compared to state-of-the-art alternatives. Visual HDR images are displayed using Reinhard tone mapping on luminance component.}
%     \label{fig:visuals}
% \end{figure*}

\section{Method}

We formulate HDR image restoration by first defining supervised learning. We then describe the input construction methodology to condition the input for the restoration model. Finally, we introduce scaling-equivariant regularization to guide learning and ensure invariance to scale operator.

\subsection{Restoration Model Formulation}

% We model the HDR reconstruction as a mapping
% \[
%     f_\theta:\;\mathbb{R}^{M\times N}\;\longrightarrow\;\mathbb{R}^{M\times N},
%     \quad
%     \hat x \;=\; f_\theta(y),
% \]
% where \(y\) is the raw modulo image and \(\theta\) denotes the learnable parameters.  The overall training objective is

We model the HDR reconstruction as the inverse mapping $\hat{\x} = f_\theta(\z)$, \bemc{where $\z$ is the input} for the network, usually the modulo image, i.e., $\z = \y$, and $\theta$ denotes the learnable parameters. The overall training objective is \begin{equation}
\label{eq:training_objective}
    \underset{\theta}{\text{minimize}}
    \;\mathbb{E}_{\z,\x }\Bigl[
      \mathcal{L}_{\mathrm{rec}}\bigl(\x,\,f_\theta(\z)\bigr)
      \;+\;\gamma\,\mathcal{R}_{\mathrm{eq}}\bigl(\z;\theta\bigr)
    \Bigr],
\end{equation}
where \(\mathcal{L}_{\mathrm{rec}}\) is a reconstruction loss that enforces consistency with HDR images and \(\mathcal{R}_{\mathrm{eq}}\) is the proposed equivariance regularizer weighted by \(\gamma>0\). \bemc{ Following classical formulation,}  we define the reconstruction loss as the minimization of the means-squared error as follows \begin{equation}
  \bemc{  \mathcal{L}_{\text{rec}}(\x, f_\theta(\z)) =  \Vert \x - f_\theta(\z) \Vert_2^2. }
\end{equation} \vspace{-2em}

\begin{figure}[!t]
    \centering
    \includegraphics[width=\linewidth]{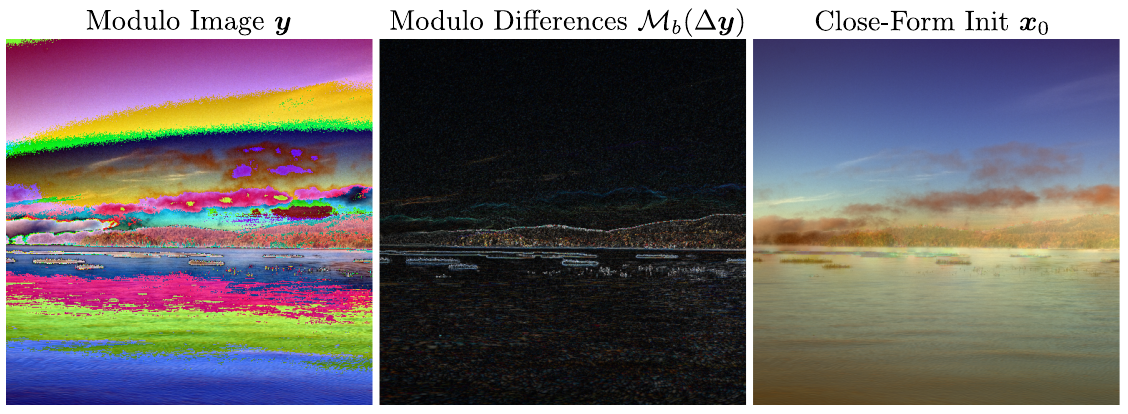} \vspace{-2em}
    \caption{\textbf{Input alternatives.} (a) raw modulo image \(\boldsymbol{y}\), exhibiting wrapping edges; (b) wrapped finite differences \(\mathcal{M}_b(\Delta\boldsymbol{y})\), which accentuate true edges; and (c) closed‑form initialization \(\boldsymbol{x}_0\), capturing large‑scale illumination.}
    \label{fig:inputs} \vspace{-1em}
\end{figure}

\subsection{Input Construction $ \z $}

We consider three distinct types of input representation for the restoration network.  In our simulations, \bemc{ each type of input is evaluated separately} and in various pairwise and full combinations to quantify how they influence each other and contribute to overall performance.

\textbf{Raw Modulo Image \( \y \).}  The primary input is the modulo image, which preserves all the content of the scene, but ``wraps" intensities in multiples of \(2^b\). Supplying \(\y\) directly ensures that the network has access to both coarse structure and fine details of the HDR scene.

\textbf{Modulo Finite Differences \(\mathcal{M}_b(\Delta \y )\).}  
To highlight edge information and correct gradient discontinuities, we calculate the modulo finite differences $\mathcal{M}_b(\Delta \y ) $.  This input explicitly allows the model to leverage accurate local gradients without having to discover modulo unwrapping solely through its learned filters.

\textbf{Closed‐Form Initialization \bemc{\(\x_0\).}} Based on 2D unwrapping problem, we derive the optimal estimation \(\ell_2\) for \(\lambda=0\)
\[
  \x_0
  = \arg\min_{\x}
    \;\bigl\lVert \Delta \x - \mathcal{M}_b(\Delta \y) \bigr\rVert_2^2,
\]
which admits a fast solution via 2D DCT~\cite{ghiglia1994robust}. By providing \(\mathbf{x}_0\), the network receives a physics‐informed starting point that captures large‐scale illumination, freeing it to focus on refining textures and correcting residual wraps.

\medskip

Although a deep network could, in principle, learn finite‐difference filters, modulo unwrapping, and DCT‐based inference from scratch, explicitly supplying \(\y\), \(\mathcal{M}_b(\Delta \y)\), and \(\x_0\) acts as a `feature lifting' strategy, \bemc{by the concatenation of different types of inputs}.  This is analogous to polynomial feature mappings, for example, augmenting \((x,y)\) with \(x^2+y^2\) to make concentric circles linearly separable~\cite{boser1992training} and allows the network to allocate its capacity to model higher-order interactions and fine-scale details.

% \subsection{Scale Equivariance Regularization $\mathcal{R}_{\text{eq}}$} 

% To learn a restoration model robust to different lighting conditions, we exploit \emph{scale equivariance} as a unitary transformation in the equivariant imaging framework.  Define the scaling operator
% \[
%     S_\alpha:\;\mathbb{R}^{m\times n}\;\to\;\mathbb{R}^{m\times n},
%     \qquad
%     S_\alpha (\x) = \alpha \cdot \x,
% \]
% where \(\alpha\) is sampled uniformly from \((a,b)\) with $a,b>0$.  We then enforce that the reconstruction process is equivariant with respect to the scale operator.  Specifically, for each training sample \( \ y\) and  \(\alpha\sim\mathcal{U}(a,b)\), we compute
% \[
%     \x_{\text{s}} = S_{\alpha}(\x),
%     \quad 
%     \y_{\text{s}} = \mathcal{W}_{b}(\x_{\text{s}})
% \]
% The equivariance regularization is
% \begin{equation}
% \label{eq:eqivariance_loss}
%     \mathcal{R}_{\mathrm{eq}}\bigl(\y;\theta\bigr)
%     = \mathbb{E}_{\alpha\sim\mathcal{U}(a,b)}
%       \bigl\lVert \x_{\text{s}} \;-\; f_\theta(\y_{\text{s}}) \rVert_2^2,
% \end{equation}
% which penalizes deviations between the scaled hdr image \( \x \) and the reconstruction.

\subsection{Scale Equivariance Regularization \(\mathcal{R}_{\mathrm{eq}}\)}

We adapt the equivariant‐imaging framework to our supervised HDR modulo restoration by treating a change in exposure as a \emph{scale} transform of the true scene.  Concretely, let \(S_\alpha:\mathbb{R}^{m\times n}\to\mathbb{R}^{m\times n}\) be defined by
\[
  S_\alpha(\boldsymbol{x}) = \alpha\,\boldsymbol{x}, 
  \quad \alpha\sim\mathcal{U}(a,b),\quad a,b>0.
\]
Given an HDR image \(\boldsymbol{x}\), we generate the scaled pair \(\boldsymbol{x}_s = S_\alpha(\boldsymbol{x})\) and its modulo measurement \(\boldsymbol{y}_s = \mathcal{W}_b(\boldsymbol{x}_s)\).  We then require the network \(f_\theta\) to satisfy the equivariance condition
\[
    f_\theta(  \underbrace{\mathcal{W}_b (S_\alpha(\x)}_{\y_{s}}) = S_\alpha ( f_\theta(\underbrace{\mathcal{W}_b(\x)}_{\boldsymbol{y}}) \approx \x_s,
\]
which we enforce via the scale equivariance loss
\[
  \mathcal{R}_{\mathrm{eq}}(\boldsymbol{y};\theta)
  = \mathbb{E}_{\alpha\sim\mathcal{U}(a,b)}
    \bigl\lVert
    \x_s - 
      f_\theta\bigl( \y_s \bigr)
    \bigr\rVert_2^2.
\]

\noindent Since changing \(\alpha\) alters the wrap‐around boundaries in \(\boldsymbol{y}_s\) without changing the underlying scene, penalizing $\mathcal{R}_\text{eq}$ teaches the network to distinguish between modulo discontinuities and natural image edges, enhancing model generalization. This supervised equivariant constraint ensures that two modulo images differing only by exposure scale share a reconstruction that is equivalent up to the scale factor \(\alpha\). \bemc{In practice, we set $a=0.9$ and $b=1.1$.}

\section{Simulations}

\subsection{Dataset and Experimental Setup}

We conduct all experiments using the UnModNet dataset~\cite{zhou2020unmodnet}, which contains paired HDR and modulo image samples specifically designed for benchmarking restoration algorithms in high dynamic range imaging. The dataset provides a controlled setting for supervised learning, enabling quantitative assessment of reconstruction fidelity under varying degrees of signal wrapping.

Performance evaluation is carried out using both perceptual and distortion-based metrics, computed in two complementary domains: the perceptually uniform PU21 encoding~\cite{azimi2021pu21} and the raw linear HDR domain. In the PU21 domain, we report PSNR computed on the luminance channel, denoted PSNR-Y, as well as SSIM-Y and MS-SSIM-Y, also measured on luminance. Additionally, we include PSNR computed over all three RGB channels to assess full-color fidelity under perceptual encoding. In the linear domain, we report PSNR and SSIM-Y over the RGB channels, referred to respectively as PSNR-L and SSIM-L. These metrics evaluate the photometric accuracy and offer a complementary perspective to the perceptual measures.

The restoration network is implemented as a lightweight variant of DRUNet~\cite{zhang2021plug}, comprising four downsampling levels with eight feature channels at the first scale. This configuration yields a total of 510.46K trainable parameters. \bemc{ Training is performed using the Adam optimizer for 5000 epochs, with an initial learning rate of \(5 \times 10^{-4}\) and a batch size of 64. The code is available open source on GitHub and is built using the DeepInvese library~\cite{monroymodulosefl, tachella2025deepinverse}.}

\begin{table}[!h]
\centering 
\caption{\textbf{Input construction influence.} The best results are \textbf{bolded}, and the second best are \underline{underlined}.} \vspace{-0.5em}
\resizebox{0.45\textwidth}{!}{
\begin{tabular}{ccc|ccc}
\toprule
\multicolumn{3}{c|}{\textbf{Input}}   &  \multicolumn{3}{c}{\textbf{Metrics on PU21}} \\ \midrule
$\boldsymbol{y}$           & $\mathcal{M}_{b}(\Delta \boldsymbol{y})$ \hspace{-1em} & $\boldsymbol{x}_0$ & PSNR-Y & SSIM-Y & MS-SSIM-Y \\ \midrule \midrule
\checkmark  &            &            & $22.95$ &  $0.800$ & $0.926$ \\
            & \checkmark &            & $19.42$ &  $0.822$ & $0.880$ \\
            &            & \checkmark & $17.00$ &  $0.744$ & $0.804$ \\
\checkmark  & \checkmark &            & $\mathbf{24.53}$ &  $\mathbf{0.853}$ & $\mathbf{0.940}$ \\
\checkmark  &            & \checkmark & $22.97$ &  $0.825$ & $0.924$ \\
\checkmark  & \checkmark & \checkmark & $\underline{23.69}$  & $\underline{0.845}$ & $\underline{0.936}$  \\ \bottomrule
\end{tabular}
}
\end{table}

\subsection{Input Construction Analysis}

Table~1 shows HDR reconstruction performance using various input configurations. The experiments include the raw modulo image \( \boldsymbol{y} \), modulo finite differences \( \mathcal{M}_b(\Delta \boldsymbol{y}) \), and closed-form initialization \( \boldsymbol{x}_0 \). Performance is evaluated using PU21 metrics: PSNR-Y, SSIM-Y, and MS-SSIM-Y.

The raw modulo image \( \boldsymbol{y} \) alone offers stable and consistent performance, serving as a solid baseline. In contrast, the wrapped finite difference input \( \mathcal{M}_b(\Delta \boldsymbol{y}) \) alone achieves moderate performance, but when combined with the modulo image, it significantly boosts restoration quality. \( \mathcal{M}_b(\Delta \boldsymbol{y}) \) provides complementary edge structure information, enhancing the network's distinction between true image edges and modulo discontinuities. \bemc{C}losed-form initialization \( \boldsymbol{x}_0 \) consistently underperforms in isolation and slightly degrades performance when combined with \( \boldsymbol{y} \) and \( \mathcal{M}_b(\Delta \boldsymbol{y}) \). Although \( \boldsymbol{x}_0 \) encodes meaningful information, residual unwrapped artifacts can interfere with the learned representation of the network. The structured prior from \( \boldsymbol{x}_0 \) is less beneficial than the network's ability to refine wrapped measurements.

We use input configuration \( \boldsymbol{z} = [\boldsymbol{y}, \mathcal{M}_b(\Delta \boldsymbol{y})] \) for all experiments due to its optimal balance of fidelity and efficiency.

\begin{table*}[!t]
  \centering
  \caption{\textbf{Quantitative results on the Unmodnet dataset.} The proposed method outperforms baselines in perceptual (PU21) and linear HDR metrics, with further gains from scale-equivariant regularization $\mathcal{R}_{\text{eq}}$.}
  \label{tab:quant_comp}
  \vspace{-0.5em}
  \resizebox{0.9\textwidth}{!}{
    \begin{tabular}{l|cccccc}
      \toprule
      \textbf{Method}       & PSNR-Y-pu & PSNR-pu & SSIM-Y-pu  & MS-SSIM-Y-pu  & PSNR-L & SSIM-L \\ 
      \midrule  \midrule
      AHFD~\cite{monroy2025autoregressive}        & \(2.64\)  & $2.55$ & \(0.468\) & \(0.510\)  & \(27.45\) & \(0.597\)  \\
      SPUD~\cite{pineda2020spud}        & \(3.00\)  & $2.94$ & \(0.489\) & \(0.539\) & \(26.75\)  & \(0.563\)    \\
      PnP-UA~\cite{bacca2024deep}      & \(3.61\)  & $3.48$  & \(0.507\) & \(0.579\)  & \(26.76\) & \(0.563\)  \\
      UnModNet~\cite{zhou2020unmodnet}    & \(20.72\) & $20.66$ & \(\textbf{0.961}\) & \(\textbf{0.968}\)  & \(28.92\) & \(0.778\)  \\
      Ours               & $\underline{24.53}$   & $\underline{22.52}$ &  $0.853$  &  $0.940$   &  $\underline{35.66}$  & \(\underline{0.970}\) \\
      Ours + $\mathcal{R}_{\text{eq}}$   & $\textbf{25.30}$   & $\textbf{23.12}$ &  $\underline{0.865}$  &  $\underline{0.948}$   &  $\textbf{36.47}$  & $\textbf{0.973}$ \\
      \bottomrule
    \end{tabular}
  }
\end{table*}

\begin{figure*}[!t]
    \centering
    \includegraphics[width=\linewidth]{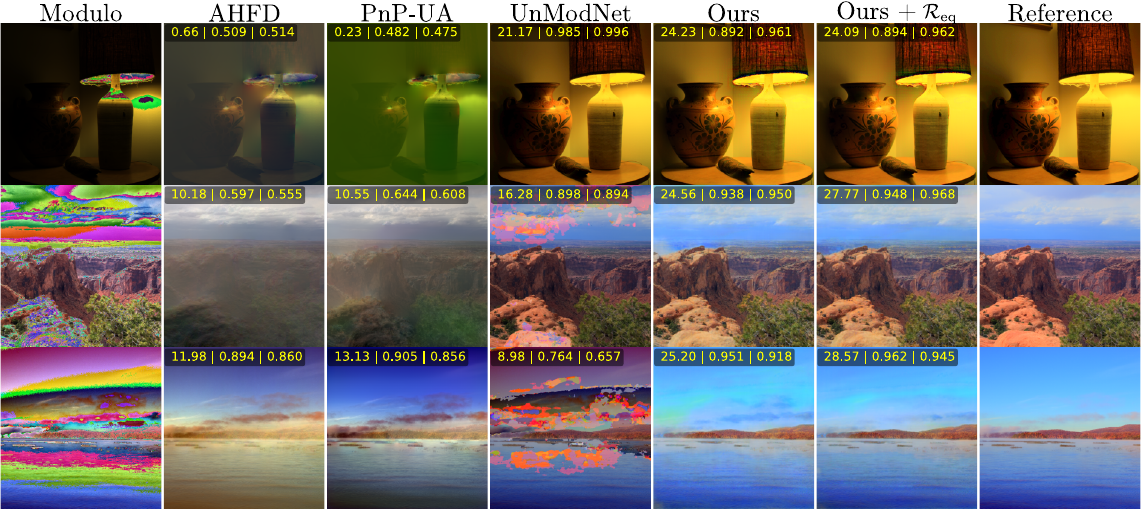} \vspace{-2em}
    \caption{\textbf{Visual Results of HDR Recovery Methods.} Qualitative comparison of unwrapped images from various methods, demonstrating the proposed approach achieves superior visual fidelity and quantitative metrics (PSNR-Y$\vert$SSIM$\vert$MS-SSIM) on PU21 encoding and compared to state-of-the-art alternatives. Visual HDR images are displayed using Reinhard tone mapping.} \vspace{-1em}
    \label{fig:visuals}
\end{figure*}

\subsection{State-of-the-art Comparison}

We evaluated the proposed method against different HDR modulo imaging restoration algorithms, in the context of modulo cameras on the UnModNet dataset. We select a deep-based unmodulo network called \textbf{UnModNet}~\cite{zhou2020unmodnet} which estimated number of folds iterative, our previous works \textbf{AHFD}~\cite{monroy2025autoregressive}, and \textbf{PnP-UA}~\cite{bacca2024deep}, which uses stripe artifact removal algorithms and denoising prior to handling the unwrapped artifact, respectively, and the simultaneous phase unwrapping and denoising (\textbf{SPUPD}) algorithm~\cite{pineda2020spud}.

Quantitative results are presented in Table~\ref{tab:quant_comp}. All methods yield comparable performance in the linear HDR domain, with PSNR-L values ranging from approximately 27dB to 36dB and SSIM-L values between 0.56 and 0.98. However, substantial differences appear when evaluating perceptual quality under the PU21 encoding. Specifically, the AHFD, SPUD, and PnP-UA methods perform significantly worse than UnModNet and the proposed approach, particularly in PSNR-Y and SSIM-Y metrics. This performance gap is due to phase-unwrapping algorithms like AHFD, SPUD, and PnP-UA producing zero-centered reconstructions, which often lack consistent offset and color alignment across channels, degrading perceptual scores in the PU21 domain. In contrast, the proposed method demonstrates superior perceptual quality, outperforming SPUD by up to 4dB in PSNR-Y and 2dB in PSNR in PU21 encoding, while maintaining competitive SSIM-Y and MS-SSIM-Y values. Furthermore, incorporating the scale-equivariant regularization term $\mathcal{R}_{\text{eq}}$  leads to additional improvements across all metrics.

\newpage

Figure~\ref{fig:visuals} presents qualitative comparisons across various HDR restoration methods. AHFD and PnP-UA struggle with highly saturated regions, such as light sources, failing to unwrap them correctly and introducing color distortions or false discontinuities. While these methods show partial success in landscape scenes, artifacts persist. UnModNet effectively unwraps concentrated highlights but fails to handle complex global variations, particularly in natural images. In contrast, the proposed method consistently reconstructs high-fidelity HDR outputs under diverse lighting conditions, achieving superior visual and quantitative performance. The addition of scale-equivariant regularization further enhances robustness, particularly in scenes with intricate luminance gradients. Despite these improvements, slight color variations are present, likely due to the wide variety of color spaces encountered in HDR imaging. These results underline the generalization capabilities of proposed method, yet indicate potential improvements in color fidelity.

\section{Conclusions}

This work introduces a learning-based framework for HDR reconstruction from modulo images, employing two complementary strategies: scale-equivariant regularization and feature lifting. The equivariant loss ensures consistency under exposure changes, enhancing the model's capability to distinguish true scene structures from artificial wrap discontinuities. Additionally, the use of lifted inputs, comprising raw modulo measurements and modulo finite differences, guides the network with knowledge priors. Quantitative results on the UnModNet dataset demonstrate that our method exceeds existing approaches by a significant margin, achieving up to 4.8dB and 2.9dB in perceptual PSNR-Y and PSNR, respectively, over UnModNet. These gains highlight the effectiveness of combining structural priors with equivariance constraints in restoring high-fidelity HDR content from challenging modulo observations.

\newpage

\bibliographystyle{IEEEbib}
\small
\bibliography{references}

\end{document}